# Parametric upconversion of Ince-Gaussian modes

HAO-RAN YANG,[1,2] HAI-JUN WU,[1,2] WEI GAO,[1] CARMELO ROSALES-GUZMÁN,[1] AND ZHI-HAN ZHU[1,*]

[1] *Wang Da-Heng Center, Heilongjiang Provincial Key Laboratory of Quantum Manipulation & Control, Harbin University of Science and Technology, Harbin 150080, China*
[2] *These authors contribute equally to this work.*
*Corresponding author: zhuzhihan@hrbust.edu.cn*

**Ince-Gaussian (IG) mode, a recently discovered type of structured Gaussian beam, corresponds to eigenfunctions of the paraxial wave equation in elliptical coordinates. This propagation-invariant mode is of significance in various domains, and in particular, its nonlinear transformation; however, there have been few relevant studies to date. In this work, we report the parametric upconversion of IG modes and associated full-field selection rule for the first time. We demonstrate that IG signals can be perfectly upconverted by a flattop-beam pump; in contrast, significant mode distortion occurred when using the most common Gaussian pump. Particular attention was given to the origin of the distortion, i.e., radial-mode degeneration induced by the sum-frequency generation excited by Gaussian pump. This proof-of-principle demonstration has great significance in relevant areas, such as high-dimension quantum frequency interfacing and upconversion imaging.** © 2019 Optical Society of America

http://dx.doi.org/10.1364/OL.99.099999

In 1961, one year after the invention of the first ruby laser [1], Franken *et al.* observed second-harmonic generation of the laser in quartz [2]. This groundbreaking demonstration commenced the field of nonlinear optics [3]. Subsequent research and further development of this field resulted in many important discoveries; in particular, parametric upconversion [4], i.e., converting a low-frequency signal to a higher frequency one, has gained significant attention because of its potential applications, such as ultraviolet-laser generation and upconversion detection/imaging [5–8]. The discovery of the orbit angular momentum (OAM) of light [9] at the end of last century renewed interest in most optical fields [10]. With regards to nonlinear optics, research on nonlinear interactions of structured light began with parametric upconversion of OAM-carrying light [11], and the transformation of OAM was then widely considered in various nonlinear systems [12–16]. Applied research in this novel field has focused on generation and manipulation of structured light (or photons) via nonlinear processes; here, the main principle (physics) relies on the transformation (or rather the selection rules) of spatial modes in nonlinear interactions.

Over the past two decades, driven by physical insights on OAM conservation, relevant research has mainly focused on the nonlinear transformation of the azimuthal structure of OAM-carrying light. Hence, Laguerre-Gaussian (LG) modes without radial indices were chosen as the object of most studies. More recently, exploration of the full field of spatial modes, which is vital for high-dimensional quantum system, has received significant attentions [17–21]. Hence, this has led to increased research interest on the nonlinear transformation of full-field LG modes, i.e., the selection rule including both OAM and radial momentum [22–24]. In this perspective, Ince-Gaussian (IG) modes are of significant interest for study [25, 26], because they are natural full-field spatial modes with well-defined Gouy phases and can be regarded as propagation-invariant superpositions of LG modes. For this, M. Krenn *et al.* systematically studied IG-mode entanglement in SPDC radiation [27, 28]. Moreover, the nonlinear transformation of IG modes involves many interesting factors, such as full-field selection rule and parity conservation. However, there have been few reports on this topic, such as ref. 29 reported generation second-harmonic IG modes by adding diffraction gratings on the crystal, and in particular, a quantitative selection rule of IG modes during nonlinear transformation is still a research gap to date.

In this letter, to fill the gap, we demonstrate the parametric upconversion of IG modes for and associated spatial-mode transformation for the first time. Here, IG signals in the near-infrared (telecom) wavelength were converted to the visible range via sum-frequency generation (SFG). Through theoretical simulations and experimental verification, we showed that the IG modes carried by the signal beams can be perfectly transferred to the SFG beam using a flattened Gaussian (or flattop) beam as the pump. In contrast, a significant distortion of the spatial modes occurred when a common Gaussian beam was used as the pump; additionally, special attention has been given to the selection rule of IG modes and how to obtain the mode composition of the distorted SFG beam.

Spatial modes refer to mathematical abstraction corresponding to the eigenfunctions of the paraxial wave equation (PWE), which enables a concise description of propagation-invariant beams. For instance, the most common LG modes are eigenfunctions with respect to cylindrical coordinates; similarly, Hermite Gaussian (HG) modes are the analogues in cartesian coordinates. Except for the fundamental TEM$_{00}$ mode, any structured Gaussian beam with a well-defined Gouy phase $(N+1)\pi/2$ [30], i.e., its intensity profile is propagation invariant, can be represented as a sum of LG or HG modes of order $N$. More interestingly, for a given order $N$, there are two complementary modes that correlate with a certain symmetry in the transverse plane, i.e., rotational symmetry (vortex handedness) for LG modes and fourfold rotational symmetry (x-y sequence) for HG modes. From this perspective, IG modes, which are eigenfunctions of the PWE in elliptical coordinates, are another complete set of spatial modes associated with parity symmetry. A pair of IG modes with even and odd parity can be expressed as:

$$\text{IG}_{N,m}^{e}(\vec{r},z;\varepsilon) = \gamma_e C_N^m(i\xi,\varepsilon) C_N^m(\eta,\varepsilon)\exp[G_N(r,z)]$$
$$\text{IG}_{N,m}^{o}(\vec{r},z;\varepsilon) = \gamma_o S_N^m(i\xi,\varepsilon) S_N^m(\eta,\varepsilon)\exp[G_N(r,z)],$$
(1)

where $\xi \in [0,\infty)$ and $\eta \in [0,2\pi)$ are the radial and the angular elliptic variables, respectively; $\varepsilon$ denotes the ellipticity; $C_N^m(\cdot)$ ($S_N^m(\cdot)$) and $\gamma_e$ ($\gamma_o$) are even (odd) Ince polynomials and the associated normalization constants, respectively; and $G_N(r,z)$ is the amplitude envelop of Gaussian beams of order $N$. It should be noted that elliptical coordinates are not unique and are determined by $\varepsilon$. Particularly, the IG modes become LG and HG modes for $\varepsilon = 0$ and $\varepsilon \to \infty$, respectively. In addition, the even and odd IG modes shown in Eq. (1) are N-order spatial modes that are *separable* states in the elliptical coordinates defined by $\varepsilon$.

For a given $\varepsilon > 0$, $\text{IG}_{N,m}^{e(o)}$ can be expressed as a superposition of $(N+1)$ LG modes with the same order $N = 2p+|\ell|$, as given by:

$$\text{IG}_{N,m}^{e(o)} = \sum_p a_p \text{LG}_p^{\pm\ell}, \quad p \in [0,1...(N-1)/2]$$
(2)

where $\text{LG}_p^{\pm\ell}$ denotes conjugate superpositions of LG modes, and $\ell$ ($p$) is the azimuthal (radial) index of the LG modes, and the complex numbers $a_p$ are modal weights. This indicates that the IG modes are full-field spatial modes in the transverse plane whose profiles are invariant upon propagation (due to having well-defined Gouy phases), and thus have excellent potential for encoding high-dimensional photonic states. Moreover, by comparing Eq. (1) and Eq. (2), we find that, in polar coordinates, IG modes are *non-separable* states with respect to azimuthal and radial indices of the LG modes.

According to Eq. (2), nonlinear transformation of IG modes usually involves complicated wave mixing of superposed LG modes. In this proof-of-principle work, we considered the practical issue of frequency upconversion of signals encoded with IG modes. Typically, except for frequency (or longitudinal mode), an ideal frequency interface that is used for upconversion detection or in communication links should not change the information carried by other degrees of freedom (DoFs) of the signal beam. In the following, we demonstrate a near ideal frequency interface for IG modes. In this work, a single-pass SFG platform was used for demonstrating the upconversion, which is given by $\kappa E^s(\omega_1)E^p(\omega_2) \to E^{up}(\omega_3)$, where $E^s$, $E^p$ and $E^{up}$ are the signal, pump, and upconversion fields, respectively, $\kappa$ denotes the coupling coefficient, and phase matching required a dispersion relation $k(\omega_1)+k(\omega_2)=k(\omega_3)$. The full-field selection rule of spatial modes during SFG is determined by the beating term $E^s(\omega_1)E^p(\omega_2)$ [24].

Figure 1 shows a diagram of the experimental setup. A 1550 nm beam collimated from a single-mode fiber (SMF) was first converted into a perfect TEM$_{00}$ mode by passing through a spatial filter. Then, it was sent to a spatial light modular (SLM-2), where it was prepared to form the target IG modes for use as the signals. For preparation of the pumps, a 795 nm beam from a SMF collimator was incident on the SLM-1, where it was converted into either flattop or Gaussian beams with variable beam sizes. For preparation of signal and pump beams, complex amplitude modulation was used [31]. The prepared pump and signal beams were combined using a dichroic mirror (DM-1), and then focused into a 20-mm-long PPKTP crystal designed for: 1550 nm + 795 nm → 525 nm (a type-0 SFG). The wavelength setting of this SFG can well separate the upconverted light (even at single-photon level) from the pump fluorescence [32, 33]. The focal lengths of L3 and L4 were both 200 mm, and the distance between L3 and the front surface of the crystal was ~178 mm. After L4, another dichroic mirror (DM-2) was used to separate the generated SFG beams (525 nm) from residual signal and pump.

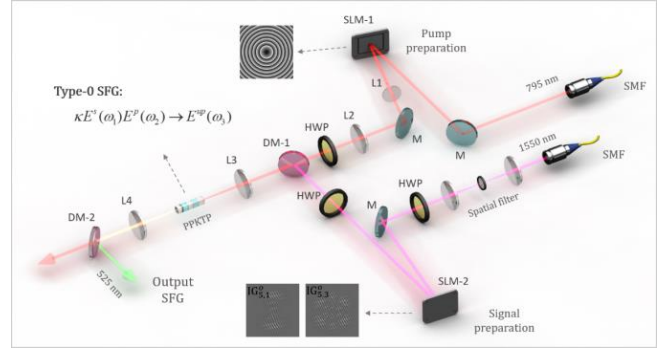

**Fig. 1.** Schematic of the experimental setup, details are provided in the text. The key components (abbreviations) include a single mode fiber (SMF), half-wave plate (HWP), mirror (M), lens (L1–L4), dichroic mirror (DM) and spatial light modular (SLM).

Without loss of generality, $\text{IG}_{5,1}^o$ and $\text{IG}_{5,3}^o$ in elliptical coordinates with $\varepsilon = 2$ were chosen as the signals. In particular, analysis of the full-field selection rule in a non-unique coordinate system is inconvenient and subject to uncertainty ($\varepsilon$-dependent). Therefore, they were represented as superposition states of the LG modes of order $N = 2p+|\ell| = 5$ for clarity, as given by:

$$\text{IG}_{5,1}^o = 0.938\text{LG}_2^{\pm 1} + 0.344\text{LG}_1^{\pm 3} + 0.048\text{LG}_0^{\pm 5}$$
$$\text{IG}_{5,3}^o = -0.343\text{LG}_2^{\pm 1} + 0.901\text{LG}_1^{\pm 3} + 0.266\text{LG}_0^{\pm 5}.$$
(3)

The IG modes shown in Eq. (3) are the full-field spatial modes in polar coordinates and are *non-separable* with respect to the azimuthal and radial indices.

It is important to note that the Gaussian (TEM$_{00}$) pump, which is commonly used for upconversion detection, does not work for these modes. This is because the nonlinear polarization (NP) excited by the rotationally-symmetric Gaussian pump is independent of the azimuthal-mode composition of the signal, but it is sensitive to the radial-mode composition. As a result, radial-mode degeneration, i.e., generating new radial modes, is introduced into the upconversion beam, leading to spatial-mode distortion. The radial-mode degeneration decreases, on the one hand, the non-separability of the azimuthal and radial modes; and on the other hand, the identity of Gouy phase of the beam.

Specifically, consider the SFG pumped by $LG_0^0$ with the same beam waist of the signal, i.e., $E^{up} \propto IG^o_{5,1(3)} LG_0^0$, according to the full-field selection rule we reported recently [22], corresponding upconversion fields can be expressed as:

$$E_{5,1}^{up} = 9.38\alpha LG_{de.}^{\pm 1} + 1.72\alpha LG_{de.}^{\pm 3} + 0.048\alpha LG_0^{\pm 5}$$
$$E_{5,3}^{up} = -3.43\beta LG_{de.}^{\pm 1} + 4.505\beta LG_{de.}^{\pm 3} + 0.266\beta LG_0^{\pm 5}, \quad (4)$$

with radial degenerated modes $LG_{de.}^{\pm 1}$ and $LG_{de.}^{\pm 3}$, given by

$$LG_{de.}^{\pm 1} = \sqrt{3/10} LG_0^{\pm 1} + \sqrt{3/5} LG_1^{\pm 1} + \sqrt{1/10} LG_2^{\pm 1}$$
$$LG_{de.}^{\pm 3} = \sqrt{4/5} LG_0^{\pm 3} + \sqrt{1/5} LG_1^{\pm 3}, \quad (5)$$

where $\alpha \approx 0.106$ and $\beta \approx 0.176$ are the normalized coefficients of intramode nonlinear coupling. This indicates that compared with the input signals, except for $LG_0^{\pm 5}$, considerable radial-mode degeneration is introduced into $LG_2^{\pm 1}$ and $LG_1^{\pm 3}$. So that the order of the spatial-modes components shown in Eqs. (5) was not unified, i.e., decreasing the identity of Gouy phase. Consequently, their beam profile would no longer be propagation invariant.

Based on Eqs. (4) and (5), images in the first line of Figs. 2(a) and (b) show the simulated beam profiles of the upconversion fields on propagation. This was confirmed by the images displayed in the second line, which shows the experimentally observed beam profiles that are in excellent agreement with the simulations. Additionally, Fig. 2(c) shows the correlation matrix of azimuthal and radial indices, where we can see that, compared with the input signals, nondiagonal elements appears in the upconversion fields. This indicates the non-separability of the azimuthal and radial modes decreased in the upconversion, which characterizes the radial-mode degeneration from another perspective.

To overcome the distortion induced by the radial-mode degeneration, we used a flattop beam that was independent to the radial mode of the input signals [34]. To obtain this beam, the most feasible way is by converting a Gaussian beam using an SLM combined with a Fourier lens [27]. Notably, the hologram (including blazed gratings) loaded on the SLM can, in principle, generating an ideal (infinite-order) super gaussian mode; but, in reality, partial high-frequency components will be inevitably filtered by the aperture of the system. In our experimental setup, the focal length of the Fourier lens (L3) was 200 mm. To optimize the overlap between pump and signal, the beam waist of the Gaussian beam to be converted was 1.4 times of that of IG signals, and the beam size at the Fourier plane was set as 0.3 mm. Based on these parameters, Fig. 3(a) shows the simulated profiles of the generated flattop beam after L3 (0–400 mm), where we can see that only a short interval (~30 mm) before 190 mm is relatively flat. It should be noted that the refractive index of the PPKTP crystal is ~1.7 time that of air, i.e., a 20 mm propagation in the crystal corresponds to a ~12 mm propagation in air. Therefore, according to the simulation, the area in the region of 178–190mm, i.e., the white dotted box plotted in Fig. 3(a), was chosen as the SFG region. Additionally, a comparison between the simulated and experimentally observed flattop beam at 190 mm is shown in Fig. 3(b), which confirmed the quality of the generated flattop beam.

To demonstrate upconversion resulting from pumping by the flattop beam, we choose $IG_{5,1}^o$ as the input signal and compare the far-field beam profiles of the signal and associated upconversion. According to the optimized flattop beam, Fig. 3(c) shows the simulated overlap of the pump and signal in the crystal (20 mm). We see that the signal is well covered by the flattop area of the pump through the crystal. The equivalent NP of the SFG (i.e., source of the SFG field) is the average of each propagation plane in the Fig. 3(c). Based on the equivalent NP, we simulated the beam profiles of the signal and upconversion light in the far field, as shown in left of Fig. 3(d). We see that this flattop-beam pumped SFG can, in principle, exactly upconvert the IG mode signals. As a proof the principle, we examined the upconversion experimentally with the optimized preferences, and the results agreed well with the simulations, as shown in the right side of Fig. 3(d).

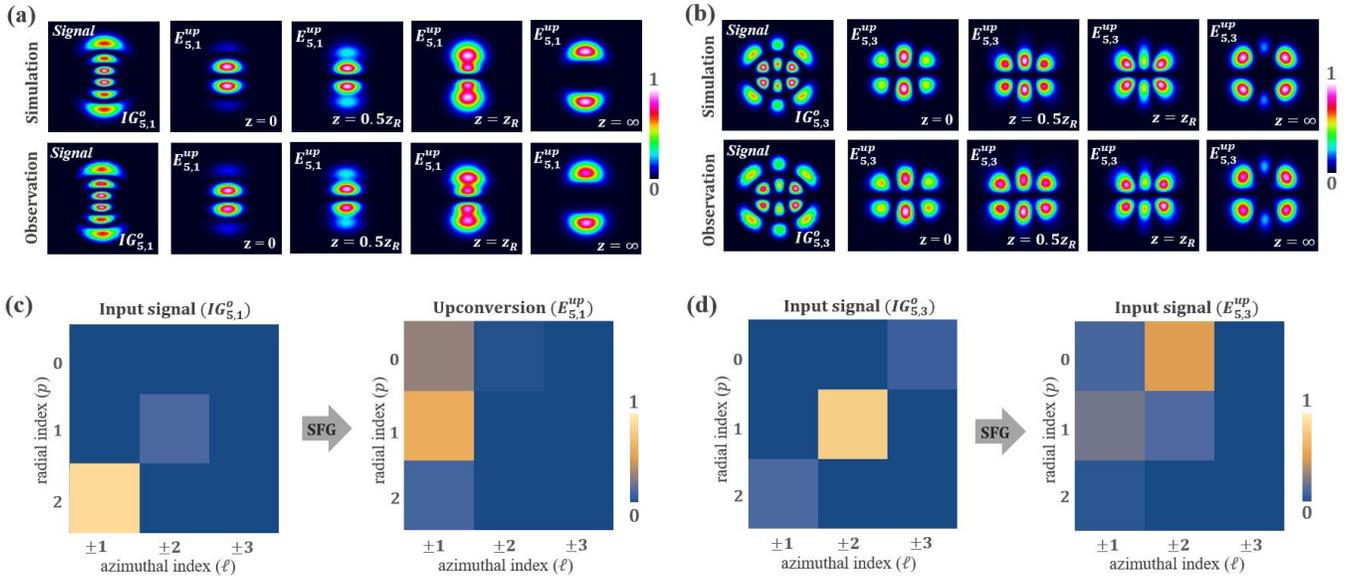

**Fig. 2.** Results of the upconversion pumped by a Gaussian beam, where (a) and (b) are the simulated and experimentally observed beam profiles of IG modes in the upconversion, respectively; (c) and (d) are corresponding correlation matrices with respect to azimuthal and radial modes.

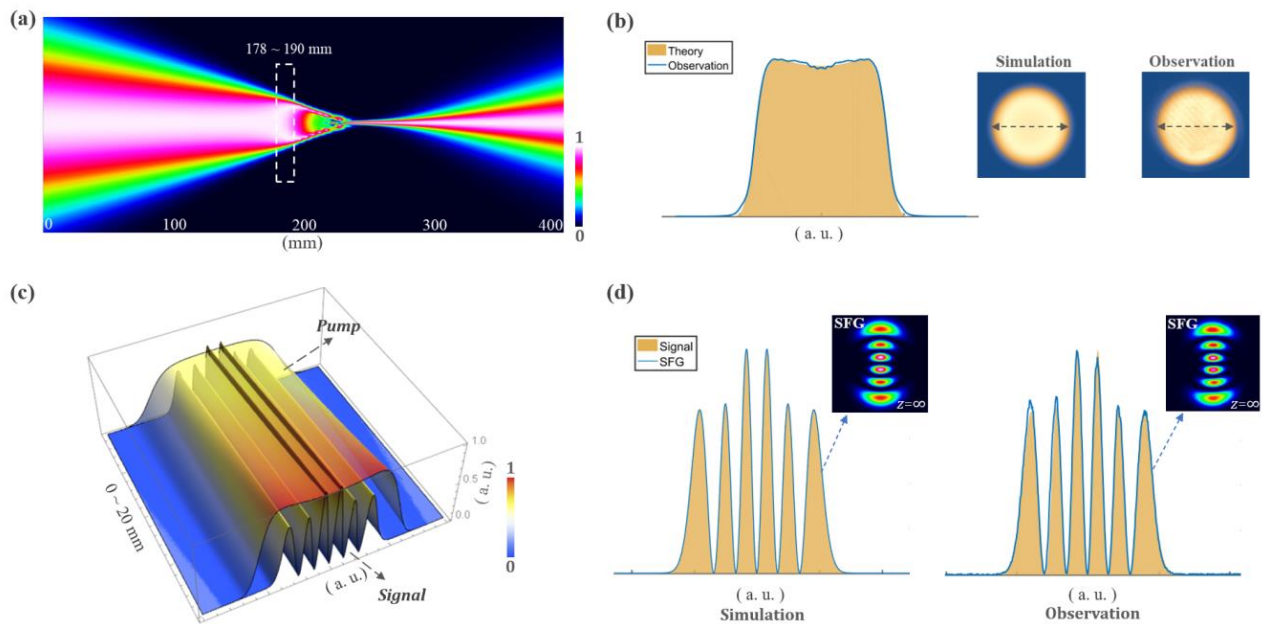

**Fig. 3**. Results of the upconversion pumped by a flattop beam. (a) Profile of the flattop beam between Lens-3 and Lens-4; (b) comparison of the simulated and observed beam profiles at $z$ = 190 mm; (c) Simulated overlap of the pump and signal in the crystal; (d) comparison of the far-field upconversion light and the input signal, obtained via simulation and experimental observation.

In summary, in this work, we studied frequency upconversion of IG modes, both theoretically and experimentally. We demonstrated that for IG-mode signals, the most commonly used Gaussian pump leads to significant mode distortion of the upconverted beams. The origin of the distortion was determined to be the radial-mode degeneration introduced during the SFG, and it was is quantitatively analyzed. To solve this thorny issue, a radially-independent SFG pumped using flattop beam was used; here, the spatial structure of the signal could be perfectly maintained in upconversion. The principle demonstrated here is of value for high-dimensional quantum frequency conversion and upconversion imaging. In particular, this upconversion technique can provide a spatial-mode-independent frequency interface for quantum frequency interface. While for the imaging, this technique is not sensitive to the location of the nonlinear device (Fourier or imaging plane) [5], which provides more freedom for designing the system. Furthermore, exploration of the parametric interaction between IG modes is of significant, and future work will address this topic.

**Funding.** National Science Foundation of China (NSFC) (11934013 and 61975047).

**Disclosures**. The authors declare no conflicts of interest.